\documentclass[aps,prl,12pt, oneside]{article}   	% use "amsart" instead of "article" for AMSLaTeX format
\usepackage[utf8]{inputenc}
\usepackage[english]{babel}

\usepackage[margin=1in]{geometry}                		% See geometry.pdf to learn the layout options. 
\usepackage[font=small]{caption}
\usepackage{graphicx}				% Use pdf, png, jpg, or eps§ with pdflatex; use eps in DVI mode
\usepackage{epstopdf}
\usepackage{amssymb}
\usepackage{amsmath}
\usepackage[table]{xcolor}
\usepackage{cite}
\usepackage[super,square,sort&compress,comma,numbers]{natbib} %superscript citation
\newcommand{\sign}{\text{sign}}

\definecolor{lightgray}{gray}{0.9}
\definecolor{pastelgreen}{rgb}{0.47, 0.87, 0.47}
\definecolor{lightapricot}{rgb}{0.99, 0.84, 0.69}
\makeatletter
\renewcommand{\maketitle}{\bgroup\setlength{\parindent}{2pt}
\begin{flushleft}
  \textbf{\@title}
\break
  \@author
\end{flushleft}\egroup
}
\makeatother

\title{{\huge Active Particles Bound by Information Flows}}

\date{}						

\begin{document}

%\twocolumn[%
     \noindent {\Large \textbf{Active Particles Bound by Information Flows}}\\
     
      \vspace{2ex}
       \noindent \large \textbf{Utsab Khadka\textsuperscript{1}, Viktor Holubec \textsuperscript{2,3}, Haw Yang\textsuperscript{1} and Frank Cichos\textsuperscript{4*}} 
       {\normalsize
       
\noindent  {\textit{$^{1}$~Department of Chemistry, Princeton University, Princeton, New Jersey 08544, USA.}}\\
       {\textit{$^{2}$~Institute for Theoretical Physics, Universit\"at Leipzig, 04103 Leipzig, Germany.}}\\
			{\textit{$^{3}$~Charles University,  
 Faculty of Mathematics and Physics, 
 Department of Macromolecular Physics, 
 V Hole{\v s}ovi{\v c}k{\' a}ch 2, 
 CZ-180~00~Praha, Czech Republic.}}\\
	   {\textit{$^{4}$~Peter Debye Institute for Soft Matter Physics, Universit\"at Leipzig, 04103 Leipzig, Germany. E-mail: cichos@physik.uni-leipzig.de}}}
\vspace{2ex}%
%]

\normalsize
\noindent \textbf{Self-organization is the generation of order out of local interactions in non-equilibrium \cite{Camazine:2001:SBS:601161}. It is deeply connected to all fields of science from physics, chemistry to biology where functional living structures self-assemble \cite{Aigouy:2010fl} and constantly evolve \cite{England:2015hl} all based on physical interactions.  The emergence of collective animal behavior \cite{Berdahl:2013bq}, of society or language are the results of self-organization processes as well though they involve abstract interactions arising from sensory inputs, information processing, storage and feedback \cite{Pearce:2014gs,Ballerini:2008cc,Attanasi:2014fc}. Resulting collective behaviors are found for example in crowds of people, flocks of birds, schools of fish or swarms of bacteria \cite{Toner:1998ke, Bialek:2012kja}. 
Here we introduce such information based interactions to the behavior of active microparticles. A real time feedback of active particle positions controls the propulsion direction these active particles. The emerging structures are bound by dissipation and reveal frustrated geometries due to confinement to two dimensions. They diffuse like passive clusters of colloids, but possess internal dynamical degrees of freedom that are determined by the feedback and the noise in the system. As the information processing in the feedback loops can be designed almost arbitrarily, new perspectives for self-organization studies involving coupled feedback systems with separate timescales, machine learning and swarm intelligence arise.}

\begin{figure}[h!]
\centerline{\includegraphics[width=0.7\columnwidth]{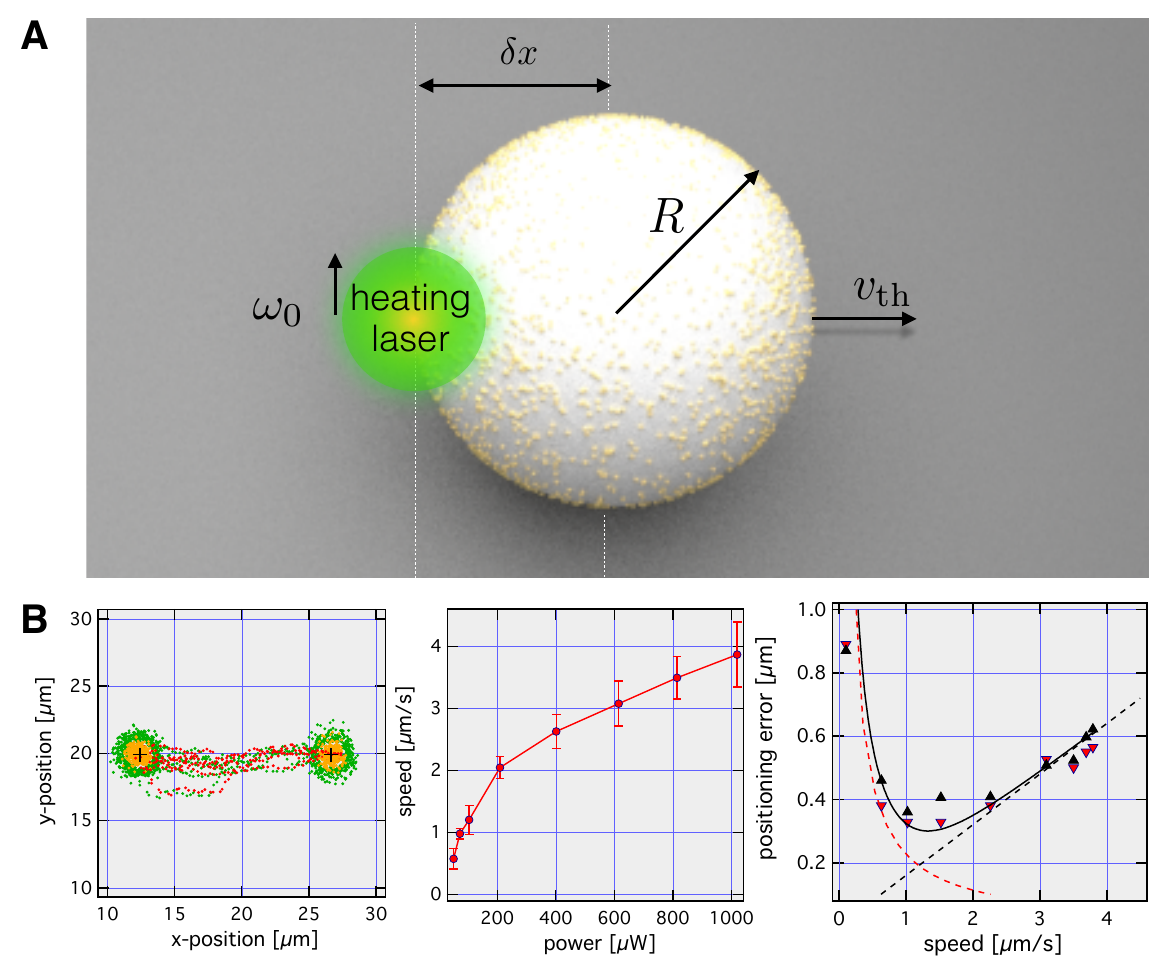}} %\textwidth
	\caption{{\bf Symmetric Self-thermophoretic Active Particles:} ({\bf A}) The self-thermophoretic active particle is composed of a melamine resin particle ($R=1.09\, {\rm \mu m}$) which is covered at 30 \% surface area by 10 nm gold nanoparticles. The nanoparticles can be heated by an incident laser beam to generate an inhomogeneous temperature profile along the particle surface. This profile causes a thermo-osmotic slip flow propelling the particle at $v_{\rm th}$ away from the laser. The velocity of the particle is determined by the displacement $d$ of the laser focus from the particle center. ({\bf B}) The particle velocity and control accuracy is derived from an experiment driving the particle between two target positions and confining it for 100 frames at each position. Example trajectory points are displayed in the left graph (laser positions (green), particle position during trapping (orange), particle positions during driving (red)).  The middle graph shows the extracted particle speed as a function of the heating power. The nonlinear dependence is analyzed in the supplementary information. The right graph displays the control accuracy as a function of the particle speed determined from the experiment in the left graph. The upwards and downwards triangles correspond to the data from the left and right target, respectively. The dashed lines are the contributions from the 2-dimensional sedimentation (red) and the particle overshooting (black). The solid black curve represents the sum of both contributions. }
\label{fig:figure1}
\end{figure}

Active particles serve as simple microscopic model systems for living objects such as birds, fish or people and mimic in particular the propulsion of bacteria or cells (without the complexity of physical properties and chemical networks in living objects). They consume energy to propel persistently and as such they have given considerable insight into collective behaviors of active materials already \cite{Palacci:2013eu,Buttinoni:2013de,Solon:2015bt,Theurkauff:2012jo,Fily:2012hj}.  With their bare function of self-propulsion they are, however, missing the important ingredients of sensing and feedback, which most living objects from cells up to whole organisms have in common. All of their living relatives have signaling inputs which they use to gain information about the environment. Using this external information, organisms such as birds and fish are able to self-organize into flocks or schools \cite{Pearce:2014gs} and, on a microscopic level, cells may regulate gene expression \cite{Tkacik:2008dq}. For flocks of birds, the structure- forming element is information and not a physical force such as Coulomb or van der Waals force. The structure formation, though, depends on the active motion of the organism and its ability to steer based on its perception of the environment \cite{Katz:2011fb, Pearce:2014gs, Swain:2012iv}. While active particles do not have such sensory inputs and feedback mechanisms built in yet, suitable control mechanisms may introduce this complexity fostering the exploration of new emergent phenomena. An information exchange between active particles has not been tackled so far, but seems to be a natural step towards extending their functionality. In this study, we demonstrate how active particles may form structures just by information exchange using a feedback control mechanism for steering the particles. 

Like bacteria, active particles have to break the time symmetry of low Reynolds number hydrodynamics in order to propel. They have to provide asymmetries to generate directed motion. The self-propulsion mechanism presented below relies on a novel scheme for generating self-thermophoresis; unlike past approaches which involved introducing the asymmetry into particle properties such as shape or material distribution, as in Janus particles, our new scheme utilizes the spatially controlled asymmetric input and release of energy around a symmetric particle. Our active particle is constructed of a melamine resin sphere with 30 \% of the surface uniformly decorated by gold nanoparticles of about 10 nm diameter. Illuminating the particle asymmetrically by a deflectable focused laser beam generates an inhomogeneous surface temperature and results in the desired self-thermophoretic motion (Figure \ref{fig:figure1} A).  
\begin{figure}[!h]
\centerline{\includegraphics[width=0.7\columnwidth]{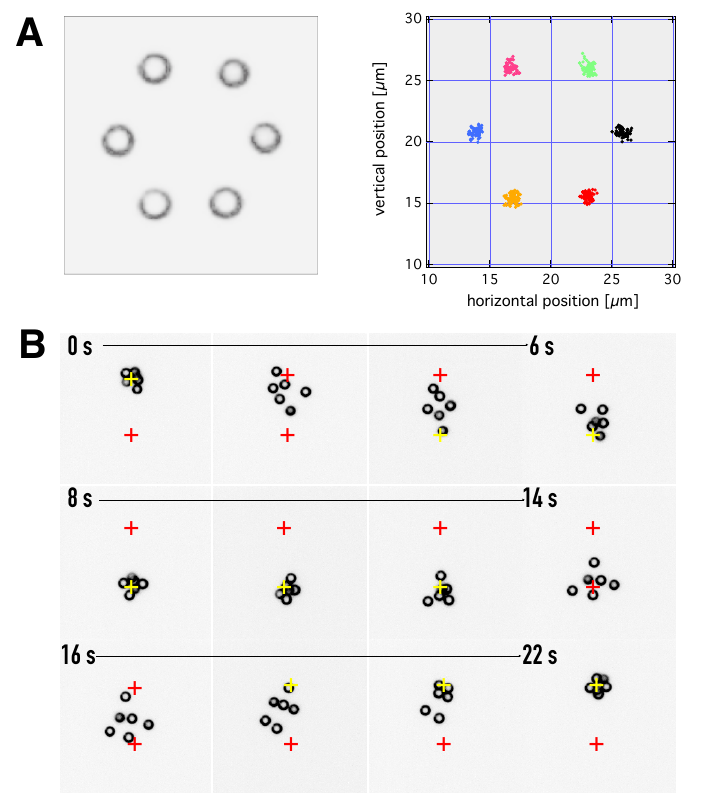}} %\textwidth
\caption{{\bf Multiple Active Particle Control.} {\bf A} Left: Darkfield microscopy snapshot (inverted grayscale) of six active particles ($R=1.09\, {\rm  \mu m}$) arranged at the nodes of a symmetric hexagon ($7.1\, {\rm \mu m}$ edge length) with the help of the particle control procedure described in the text. The incident laser power per particle is $P=0.2\, {\rm mW}$. Right: Corresponding trajectory points of the particles over a time period of $11\, {\rm s}$ with $\Delta t_{\rm exp}=110\, {\rm ms}$ exposure time/inverse frame rate.  {\bf B} Dark field microscopy image series (inverted grayscale) of six active particles driven between two target positions. Targets are colored for clarity. The incident heating power per particle is $P=0.2\, { \rm m W}$ and the time resolution of the experiment is $\Delta t_{\rm exp}=80\, {\rm ms}$. }%(Movie_16-08-09_200234)
\label{fig:figure2}
\end{figure}
This design allows for a new control scheme for active particle steering. To control the particle propulsion direction, we placed the laser beam's focal spot near the circumference of the particle. The propulsion direction is then the vector from that heated circumferential spot to the particle center. Different from standard active particles, the timescale of the rotational diffusion of the particle is irrelevant due to the missing particle asymmetry. This scheme delivers a precise control of each individual particle in a larger ensemble, a basic requisite for self-organization experiments presented here.

We first evaluate properties of a single active particle and the control accuracy in an experiment driving the particle between two target positions and confining it for a certain time at the targets (see left panel of Figure 1 B).
\begin{figure}[!h]
\centerline{\includegraphics[width=0.7\columnwidth]{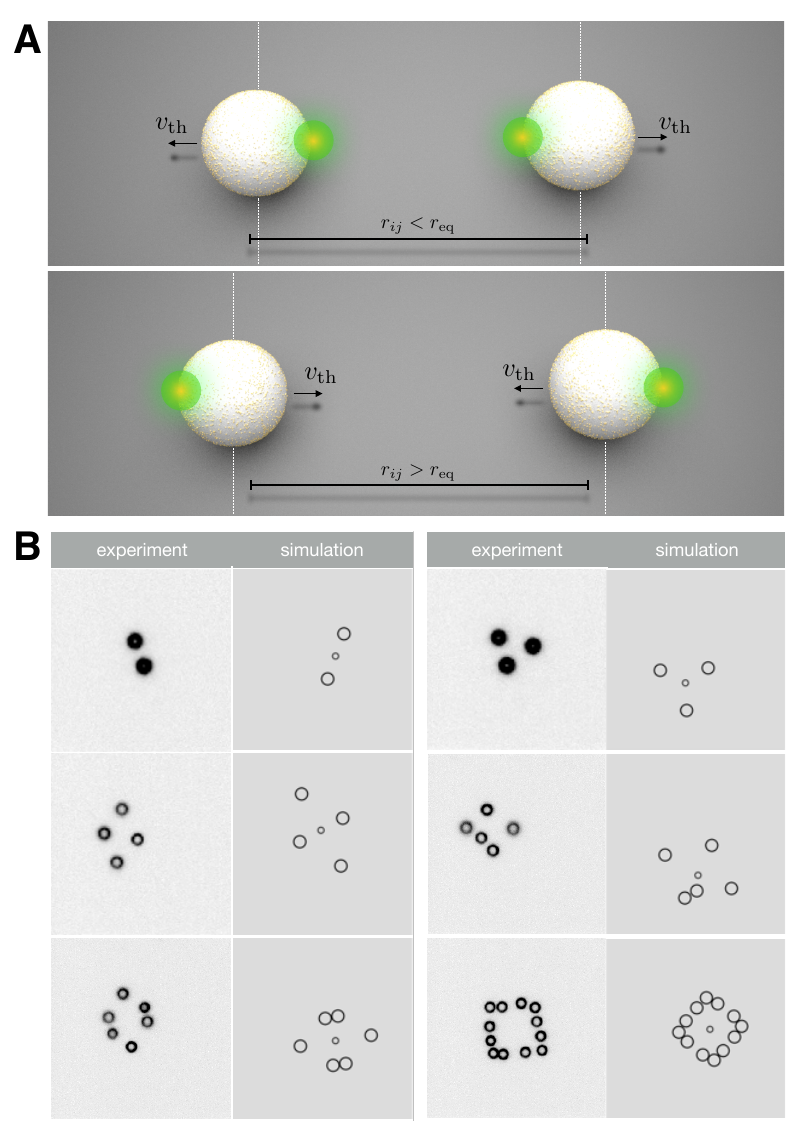}} %\textwidth
\caption{{\bf Self-organized Active Particle Molecules.} {\bf A} Pair interaction rule used to "bind" the active particles to each other. Particles are propelled with a speed $v_{\rm th}$. If the distance is below $r_{\rm eq}$, the particle is moved away from the other, if above, they are propelled towards each other. {\bf B} Example snapshots of 6 different "active particle molecules" that are bound by the interaction rule in A. The structures are highly dynamic (see sample movies in the SI). The experimental images are compared to snapshots of corresponding numerical simulations involving Brownian motion and a simple delayed feedback as in the experiment. The smaller central circle marks the center of mass. An effective potential description also reveals equivalent structures up to the pentamer. }
\label{fig:figure3}
\end{figure}
We extract the dependence of the active particle propulsion velocity on the heating power from this experiment finding a non-linear scaling (middle panel Figure \ref{fig:figure1} B) as the particle slips away from the focus during the camera exposure time. Incorporating this slipping process in a model reproduces this nonlinear increase qualitatively (see SI). The confinement at the target position can be characterized by a positioning error $\sigma=\sqrt{\langle {( \bf r} - {\bf r_{t}})^2\rangle}$ -- the standard deviation of the probability density of the particle--target distance in the steady state.  Here, $\bf r$ and ${\bf r}_t$ are the coordinates of the swimmer and target, respectively. The positioning error  obeys two regimes \cite{Selmke:2017gya,Selmke:2017gy,Bregulla2014,Qian:2013eha}. When the displacement of the particle due to the propulsion is smaller than the diffusive displacement during the exposure time $\Delta t_{\rm exp}$, the positioning error is reflected by a simple sedimentation model. A constant particle speed $v_{\rm th}$ drives the particle radially  towards the target position against the Brownian motion with a diffusion coefficient $D_{0}$. Hence, the density distribution in the steady state is exponential with a characteristic length scale (the sedimentation length in two dimensions) $\rho = \sqrt{6} D_{0}/v_{\rm th}$ as indicated by the red dashed curve in the right panel of Figure \ref{fig:figure1} B \cite{Selmke:2017gy}. When the active particle speed increases, an overshooting of the particle over the target position due to the finite sampling of the position of the particle with the camera exposure time $\Delta t_{\rm exp}$ defines the positioning error \cite{Jun:2012bx}. The overshooting distance equals the traveled distance within the time between two frames $\Delta t_{\rm exp}$ and increases linearly with the velocity of the active particle as shown by the black dashed line. The sum of both contributions determines the positioning error $\sigma$ which is depicted for different particle speeds from the experiments (triangle markers, forward and backward motion in the left graph of Figure \ref{fig:figure1} D) together with the predicted curve $\sigma=\sqrt{6 D_{0}^2/c^2 v_{\rm th}^2+c^2 v_{\rm th}^2\,\Delta t_{\rm exp}^2}$ with no free parameters (solid black curve). A minimum positioning error of $\sigma=370\, {\rm nm}$ is found for the exposure time of $\Delta t_{\rm exp}=80\, \rm ms$, a diffusion coefficient of $D_{0}=0.23\,  \rm \mu m^{2}/s$ (experimentally found as compared to theory $D_{0}=0.20\,  \rm \mu m^{2}/s$) and a velocity of $v_{th}=1.3\, \rm \mu m /s$. A factor $c=2$ accounts for the fact that the particle travels twice the distance due to the feedback delay of one exposure time $\Delta t_{\rm exp}$.
%To establish virtual interactions between many active particles, the control scheme requires an independent steering of multiple particles. This 
\paragraph{Multiple Active Particles, Swarms and Structures}
 Multiple particle control is introduced by illuminating multiple particles at suitable positions at their respective circumferences. In the current setup, an acousto-optical deflector multiplexes the focused heating laser spot between different particle positions within one exposure of duration $\Delta t_{\rm exp}$ (see methods section). The incident heating power is therefore available for a time $\Delta t_{\rm exp}/N$ to each of the $N$ particles and the average heating power per particle decreases when keeping the overall incident laser power constant. 
Figure \ref{fig:figure2} A depicts the control of six individual active particles in a spatially fixed pattern of six target positions, arranged as the nodes of a symmetric hexagon. The particles were initially distributed randomly in the field of view. Once the control was initiated, each of the particles was first driven towards its nearest target, after which it was confined there. The resulting steady state distribution of the particles overlaps with the distribution of the assigned target coordinates, as shown in the Figure. The accuracy of the particle control follows the dependencies revealed earlier in Figure \ref{fig:figure1}. As compared to optical tweezers, the applied scheme does not involve external forces, but just dissipative fluxes and is thus physically different from the common trapping\cite{Qian:2013eha}. Yet, the stationary position distribution of each particle around its target may be converted into a virtual effective potential \cite{Jun:2012bx, Braun:2015dla}. 

A collection of particles may be driven to a target either by first arranging them into a structure and subsequently translocating the structure, or by driving each particle directly to the target without prior structural arrangement, as demonstrated in the supplementary movies. While the former approach results in a collective motion resembling the transport of an organized fleet of vehicles (see sample movies in the SI), in the latter case, the unstructured collective driving results in a swarm-like motion determined by the active particles steric repulsion, Brownian motion and the propulsion towards the target (see Figure \ref{fig:figure2} B). In addition to the switching of propulsion directions, the local modulation of the propulsion speed may also lead to a well-controlled effective potential sculpting the particle probability density distributions according to $p({\bf r})\propto 1/v_{\rm th}({\bf r})$ much like in the motility-induced phase transitions observed in active particle ensembles \cite{Cates:2013ia, Cates:2012is}.

\paragraph{Self-organized Active Particle Molecules}
The described multiple particle control scheme opens the realm of feedback-induced virtual interactions for active particles. Here, particles are allowed to exchange information via the real-time tracking feedback loop of the microscopy system. It is similar to a situation where birds or fish react to the action of their neighbors to form flocks and schools or to escape predators. It introduces a signaling channel between the particles, which can be tweaked almost arbitrarily to design virtual interactions and paves the way for a vast amount of studies from the self-organization of new structures to the information flow in flocks \cite{Attanasi:2014fc}, or the application of machine learning to study adaption and the emergence of collective patterns \cite{Palmer:2017up}. Moreover, it provides a minimal scalable robotic system with a simple propulsion and intrinsic noise due to Brownian motion.

Here, we demonstrate the structure formation by defining a pairwise control, which intends to keep the active particles at a prescribed separation distance $r_{\rm eq}$ by just changing their propulsion direction but not the speed. If the in-plane distance $r_{ij}$ between two particles ($i$ and $j$) is below the separation distance $r_{\rm eq}$, the particles are pushed away from each other, each with a speed $v_{\rm th}$. In the case $r_{ij}>r_{\rm eq}$, the particles are pushed towards each other with the same speed, which results in an effective V-shaped interaction potential for a pair of active particles. For a number of $N$ interacting particles, this feedback rule is represented for particle $i$ by the velocity
\begin{equation}\label{eq:pair}
{\bf v}_{i}(t)=-v_{\rm th}\,{\bf e}_{i}(t),
\end{equation}
where the propulsion direction is determined by
\begin{equation}\label{eq:pair_uv}
{\bf e}_{i}(t)=\frac{\sum\limits_{j\neq i}^N  \sign(r_{ij}(t-\delta t)-r_{\rm eq}){\bf e}_{ij} }{|\sum\limits_{j\neq i}^N  \sign(r_{ij}(t-\delta t)-r_{\rm eq}){\bf e}_{ij} |}	
\end{equation}
with $r_{ij}=|{\bf r}_{j}(t-\delta t)-{\bf r}_{i}(t-\delta t)|$ and ${\bf e}_{ij}=({\bf r}_{j}(t-\delta t)-{\bf r}_{i}(t-\delta t))/r_{ij}$. As stated by equation \ref{eq:pair}, the speed of motion is always $v_{\rm th}$, while its direction ${\bf e}_{i}(t)$ is defined by the positions of the other active particles. 

The interaction rule is implemented in a real time particle tracking loop (see methods) detecting the center positions of the active particles and steering the heating laser to the corresponding spots at their circumferences. Position measurements and action are separated by one exposure time $\Delta t_{\rm exp}$ introducing the feedback delay $\delta t=\Delta t_{\rm exp}$. 

\begin{figure*}[!h]
	 \centerline{\includegraphics[width=0.9\textwidth]{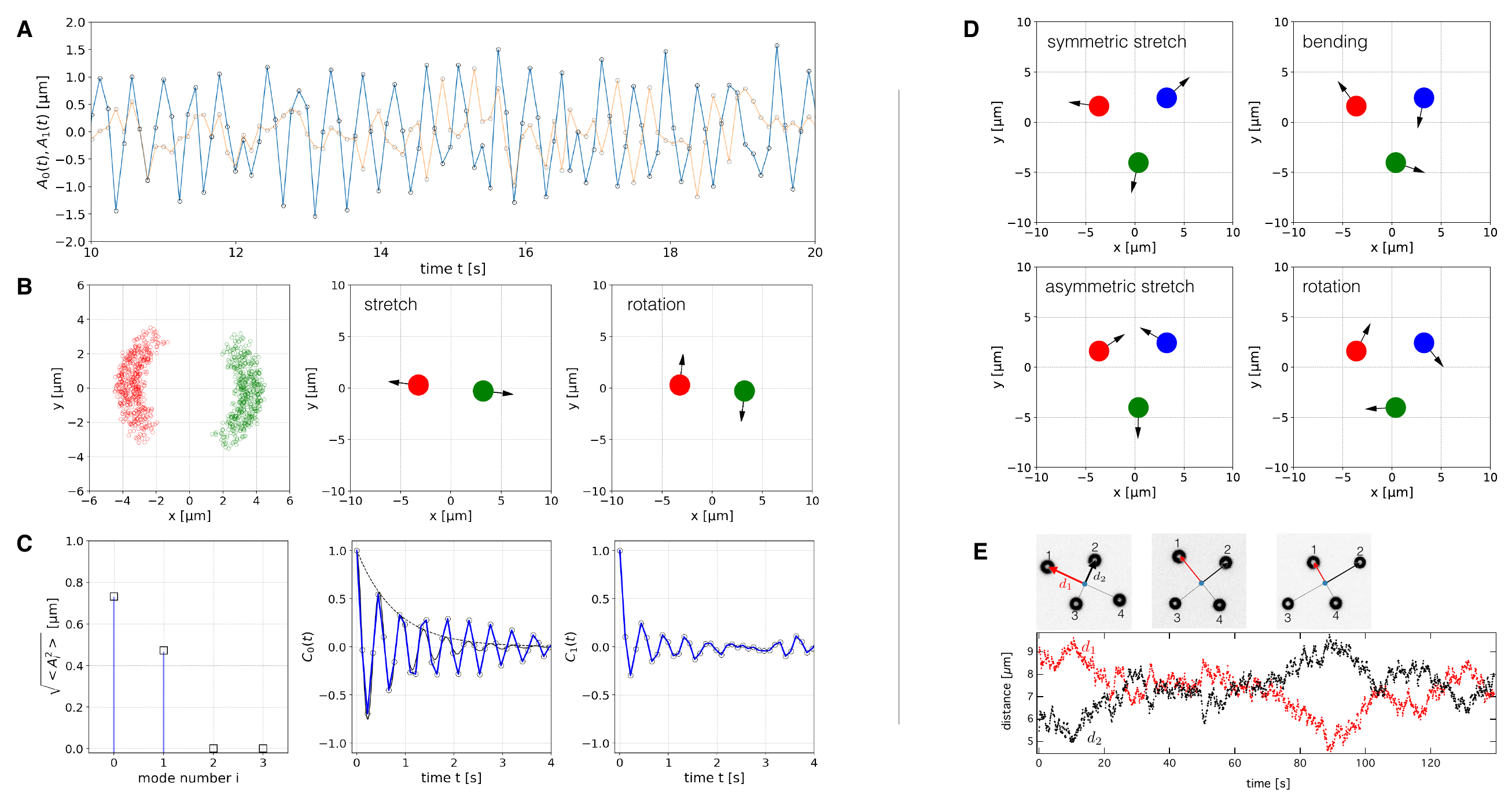}} %\textwidth
   	\caption{{\bf Active Particle Molecule Dynamics:} {\bf A} Change of the dimer bond length parallel (blue) and perpendicular (orange) to the connecting line of the two particles. The dimer bond length is oscillating with a triangular shaped elongation with a period of $T=0.44\, {\rm s}$. The period corresponds to four times the feedback delay time $\delta t=0.11\, {\rm s}$. {\bf B} Left: Trajectory points of the two bound particles in the center of mass (COM) frame. Middle and Right: Principle components of the particle displacements in the center of mass frame as obtained from a principle component analysis (PCA). {\bf C} Left: Principle component amplitudes as calculated for the two modes in B. Middle and Right: Autocorrelation functions for the displacements of the particles in the center of mass frame of the two eigenvectors obtained from the PCA. The two modes reveal a damped oscillation due to the Brownian motion of the active particles. The middle graph shows in addition the theoretical prediction for the oscillation (black solid line) and the exponential decay due to the dephasing (black dashed line). {\bf D} Principle components of the trimer are a symmetric stretch (top left), bending motion (top right), asymmetric stretch (bottom left) and a rotation (bottom right) as obtained from the experimental data. {\bf E} Isomerization of the tetramer. Top images represent the observed darkfield images of the active particles, while the lower graph displays the two marked center of mass distances of the two marked particles as functions of time. The isomerization is due to Brownian motion and occurs on timescales of $100\, {\rm s}$. }
	\label{fig:figure4}
\end{figure*}

After a short self-organization phase, the particles arrange into dynamic structures that are reminiscent of simple molecules (see SI). Figure \ref{fig:figure3} depicts snapshots of 6 self-organized structures all with $r_{\rm eq}=7.1\, {\rm \mu m}$. Active particle molecules with $N=2$ and $N=3$ particles (dimer and trimer) form structures where all average interparticle distances correspond to the adjusted value of $r_{\rm eq}$ (measured $\langle r_{ij} \rangle=7.23\, {\rm \mu m}$). All clusters with a larger number of particles ($N\geq 4$) are structurally frustrated due to the confinement in two dimensions. As a result, 4 active particles cannot form a tetrahedral structure with equilateral triangular faces. Instead a 2-dimensional structure with two "isomers" is found (Figure \ref{fig:figure4} E, shown together with a timetrace of the isomerization process). 
Larger active particle molecules ($N>4$) form frustrated structures with interparticle separations of less than the defined value $r_{\rm eq}$. Figure \ref{fig:figure3} highlights snapshots of a pentamer, a hexamer and a dodecamer. The dodecamer, for example, forms a transient square structure out of 4 right angle subunits of 3 particles. 

The structures are formed due to the information exchange between sample and feedback loop. At a distance of 7.1 ${\rm \mu m}$ hydrodynamic, electrostatic or thermal interactions are negligible and there is no spatial variation of propulsion speed, i.e. due to a mutual locking of particles which is relevant for the dynamic cluster formation in active particle suspensions \cite{Palacci:2010hk, Buttinoni:2013de}. Yet, there is a continuous entropy production in the system which comprises different contributions. The first contribution is a constant entropy production rate maintaining the temperature gradients and propelling the particles, which is required for the mobility but not sufficient for the structure formation. The emerging artificial molecules are the result of the position measurement and feedback extracting entropy from the system by steering the propulsion direction. Besides the measurement and feedback actions there is a continuous loss of structure due to Brownian motion. To form a stable stationary structure, the entropy extracted per time unit in the feedback loop has to compensate at least the increase of entropy per time unit due to Brownian motion. The structure formation is thus the result of the information flow in the feedback loop only. 

A detailed picture on the experimentally observed dynamics of the active particle structures is obtained from a principle component analysis (PCA) of the displacement vectors of the individual particles between successive frames in the center of mass frame $\Delta \vec{r}_{i}^{\,\rm COM}$. The identified eigenvectors correspond to a set of $2N-2$ orthogonal directions with the largest displacement variances in the COM frame. Figure \ref{fig:figure4} B and D present the two modes for the dimer structure (stretch and rotation, $N=2$) and the four modes of the trimer (symmetric stretch, bending mode, asymmetric stretch and rotation, $N=3$) where the modes agree well with the normal modes of an equilateral triangular structure in two dimensions. The appearance of a rotational mode is at first glance surprising as the feedback is designed to act along the connecting line between the bound particles. Due to the feedback delay and Brownian motion, the laser heating position along the circumference is fluctuating as well and introducing a coupling between the translational and rotational motion. 
The oscillatory motion as indicated for the two modes of the dimer (Figure \ref{fig:figure4} A) is a fundamental feature of a time--delayed negative feedback system and inherent to electronic oscillators, but also appearing at all levels of biological systems. In our active particle assemblies the dynamics is controlled by three parameters, the feedback delay $\delta t$, the propulsion velocity $v_{\rm th}$ and the single particle diffusion coefficient $D_{0}$. Using these parameters and restricting the analysis to the dynamics of the dimer along the connecting line (stretch mode), we can model the stretch mode dynamics with an overdamped Langevin description,

\begin{equation}\label{eq:motion}
	\dot{r}_{12}(t)=-2 v_{\rm th}\, \sign\left ( r_{12}(t-\delta t)-r_{\rm eq}\right )+\sqrt{4D_{0}}\eta_{12}(t).
\end{equation}

Here $\eta_{12}(t)$ is a zero-mean, unit-variance Gaussian white noise, i.e. ($\langle\eta_{12}(t)\rangle = 0$ and $\langle\eta(t)\eta(t')\rangle = \delta(t-t')$) such that the variance of the noise term in equation \ref{eq:motion} corresponds to $4D_{0}$. As the bond length involves the relative motion of two particles, the relative velocity $2 v_{\rm th}$ and the relative diffusion coefficient $2D_{0}$ enter the equation. Its solution yields the observed oscillatory motion with a triangular shape and an oscillation period of $T=4\delta t$ (see SI). Accordingly, the oscillatory dynamics vanishes for zero feedback delay and stiff structures appear. The amplitude of the motion linearly depends on the delay and the active particle velocity ($\Delta r_{12}=2 v_{\rm th}\delta t$).  A feedback delay of $\delta t=0.11\, {\rm s}$ and a propulsion velocity $v_{\rm th}=3.4\, {\rm \mu m/s}$ ($P_{\rm heat}=0.75\, {\rm mW}$ per particle) delivers $T=0.44\, {\rm s}$ and $\Delta r_{12}=0.75\, {\rm \mu m}$, which compares well to the experimental data for the dimer $T_{\rm exp}=0.44\, {\rm s}$ and the amplitude of the stretch mode $\sqrt{<A_{0}^2>}=0.72\, \mu {\rm m}$ obtained from the first eigenvalue of the PCA, $A_{0}$. The effect of the last term in equation \ref{eq:motion} is to introduce phase and amplitude noise to the oscillatory motion. The oscillations are therefore losing coherence and the autocorrelation $C_i(t)=\langle A_i(\tau)A_i(\tau+t)\rangle_{\tau}/\langle A_i(\tau)^2 \rangle_{\tau}$ of the oscillating modes $A_{i}(t)$ decays. The timescale of this damping is the dephasing time, which is termed $T_{2}$ in molecular spectroscopy. Using an approximate solution of equation \ref{eq:motion} for the dimer (see SI), we find a dephasing time $T_{2}\approx 32 \Delta r_{12}^2 \pi^{-4} D_{0}^{-1}$, which scales inversely with the strength of the noise given by the diffusion coefficient (see SI).  For the dimer bond length oscillation displayed in Figure \ref{fig:figure4} C we find a dephasing time of $T_{2}=0.8\, \rm s $ (dashed line). In larger structures, each particle contributes to the total noise such that coherent oscillations as observed for the dimer disappear quickly with growing size of the cluster. 

Propulsion speed, Brownian motion as well as the feedback related information flow shape the morphology and dynamics of these artificial self-organized active structures. Much like for their macroscopic counterparts, they require no external forces, but reveal features of systems with physical interactions though it is the feedback delay which controls their dynamics. Using the described method, almost any type of interaction can be designed to create large scale interacting assemblies or new self-organized shapes and which may not be accessible by conventional interactions. %With this flexibility one can readily create, for instance, the first experimental Vicsek type swarming of active particles. 
Fundamental interaction rules for emergent complex behavioral modes may be explored employing machine learning algorithms including predictive information or reinforcement learning. The details of information flows in large ensembles may be studied easily and can be connected to different timescales of delayed information processing. Especially the latter type of application including coupled active feedback networks with different inherent timescales shall ignite a vast variety of research on emergent collective and crowd dynamics.

\paragraph{Acknowledgement}

Discussions with J. Shaevitz (Princeton University), K. Kroy (Universit\"at Leipzig) and help with the sample preparations by D. Cichos (Berlin) are acknowledged. HY and UK acknowledge support by the Betty and Gordon Moore foundation (grant \# 4741). VH is supported by a Humboldt grant of the Alexander von Humboldt Foundation and by the Czech Science Foundation (project No. 17-06716S). FC is acknowledging support by grant CI 33/16-1 of the German Research Foundation (DFG).

\section*{Materials}
Samples consist of commercially available gold nanoparticle coated melamine resin particles of a diameter of 2.13 ${\rm \mu m}$ (microParticles GmbH Potsdam, Germany). The gold nanoparticles are covering about 30 \% of the surface and are between 8 and 30 nm in diameter. Glass cover slips have been dipped into a 5 \% Pluronic F127 solution, rinsed with deionized water and dried with nitrogen. The Pluronic F127 coating prevents sticking of the particles to the glass cover slides. Two micro-liters of particle suspension are placed on the glass cover slides to spread about an area of 1 cm $\times$ 1 cm to form a 3 ${\rm \mu m}$ thin water film. The edges of the sample have been sealed with silicone oil to prevent water evaporation.

\section*{Methods}
\paragraph{Microscopy Setup}
Samples have been investigated in a custom-built inverted microscopy setup (see Supplementary Information). The setup is based on an Olympus IX 71 microscopy stand. Optical heating of the active particles is carried out by a CW 532 nm laser. The laser intensity is controlled by a Conoptics 350-50 electro-optical modulator. An acousto-optic deflector (AOD) together with a 4-f system (two $f=20\, {\rm cm}$ lenses) is used to steer the 532 nm wavelength laser focus in the sample plane. The AOD is controlled by an FPGA (National Instruments) via a LabView program. The calibration of the AOD for precise laser positioning is carried out using a 2D projection method. A Leica 100x, infinity-corrected, NA 1.4-0.7 (set to 0.7), HCX PL APO objective lens is used for focusing the 532 nm laser to the sample plane as well as for imaging the active particles. Active particles are imaged under dark field illumination using an oil immersion dark-field condenser. The scattered light from the sample is collected with the Leica objective lens and imaged with a $f=30\, {\rm cm}$ tube lens to an emCCD camera (Cascade 650). A region of interest (ROI) of 200 x 200 pixels is utilized for the real time imaging, analysis and recording of the particles, with an exposure time $\Delta t_{\rm exp}$ of 0.08 s or 0.110 s. 

\paragraph{Single Particle Tracking and Real-Time Feedback Loop}
The active particles appear as rings in dark-field microscopy images and are tracked in real time in a LabView program. A Matlab node in the LabView determines the centers of the particles using a Hough transform function of Matlab. The particle coordinates are used to calculate the position of the laser focus for each individual particle. During one camera exposure of $\Delta t_{\rm exp}$, the laser is shared among the particles selected for feedback control. The switching is done using the AOD as mentioned above.

For the control experiments (Figure \ref{fig:figure1}), the controlled particle is driven back-and-forth between two prescribed target positions. Upon reaching a target, it is actively positioned there for 100 frames first, before being driven to the the target. This procedure is repeated several times (5--7) with different laser powers up to 1  mW. The particle positions around the targets are used to determine the localization error $\sigma=\sqrt{<\delta {\bf r}^2>}$, where $\delta {\bf r}={\bf r}-{\bf r}_{t}$ is the 2-d position vector from the target to the particle. Velocities are determined by projecting the particle displacement between two subsequent frames onto the unit vector given by the laser position and the particle center $\langle v \rangle=\langle ({\bf r}(t+\Delta t_{\rm exp})-{\bf r}(t))\cdot {\bf e}_{\rm lp}\rangle/\Delta t_{\rm exp}$, where ${\bf e}_{\rm lp}=({\bf r}-{\bf r}_{\rm laser})/|{\bf r}-{\bf r}_{\rm laser}|$. 

\paragraph{Principle Component Analysis of the Active Particle Molecule Dynamics}

The experiments sample the position vector $\mathbf{r}_i(t)$ for each of the $N$ particles in $N_{\rm steps} = t_{\rm meas}/\Delta t_{\rm exp} + 1$ time instants. The particle coordinates are converted into the center of mass frame of the structure. In two dimensions, we obtain $N_{\rm data} = 2\times N \times N_{\rm steps}$ data points, which we analyze using the principal component analysis. 

First, we construct the $2 N$ time-series of displacements of the individual degrees of freedom in our experiment (coordinates of the individual $N$ particles). Second, we put all these displacements for a given $t$ into a single vector: $\mathbf{S}(t) = (\Delta x_1(t),\Delta y_1(t),\dots,\Delta x_N(t),\Delta y_N(t))$. Then, we construct the matrix $\mathbf{X}$ containing in its lines the vectors $\mathbf{S}(t)$ corresponding to the individual measurement times, so the element $[i,j]$ of this matrix is given by $X_{ij} = S_j((i-1)\Delta t_{\rm exp})$. The matrix $\mathbf{X}$ thus has $2N$ columns and $N_{\rm steps} - 1$ rows. The matrix $\mathbf{X}$ is used to calculate the covariance matrix $\mathbf{M} = \mathbf{X}^T \mathbf{X}$. This matrix $\mathbf{M}$ is a symmetric $2N\times 2N$ matrix with the elements $M_{ij} = \sum_{k=1}^{N_{\rm steps} - 1} X_{ki}X_{kj} = \sum_{k=1}^{N_{\rm steps} - 1} S_i((k-1)\Delta t_{\rm exp}) S_j((k-1)\Delta t_{\rm exp})$. The diagonal thus contains the variances corresponding to the degrees of freedom measured in the experiment, i.e. of the displacements of the positions of the individual particles.

We determine all $2N - 2$ nonzero eigenvalues $A_i$ and
normalized eigenvectors $\mathbf{V}_{i}$ of the matrix $\mathbf{M}$.
%We determine all $2N-2$ eigenvalues $A_i$ and normalized eigenvectors $\mathbf{V}_{i}$ of the matrix $\mathbf{M}$. 
The individual eigenvectors determine new $2N-2$ collective degrees of freedom, which are mutually independent. For example for the dimer, the eigenvector with the largest eigenvalue determines the vibrational mode and the corresponding collective coordinate is proportional to the vector connecting the two particles. 

The vector form of the time series corresponding to the mode given by the column eigenvector $\mathbf{V}_i$ can be obtained by projecting the eigenvector $\mathbf{V}_{i}$  onto the matrix $\mathbf{X}$, i.e. $\mathbf{K}_i = \mathbf{X} \cdot \mathbf{V}_i$. The elements of the $\mathbf{K}_i$ represent the time series $A_i(t)$ of the motion along the eigenvector $\mathbf{V}_i$. To access the dynamics of the mode we calculate its autocorrelation $C_i(t)=\langle A_i(\tau)A_i(\tau+t)\rangle_{\tau}/\langle A_i(\tau)^2 \rangle_{\tau} $. The subscript t denotes that the correlation function is obtained from the time average (see SI).
 
%\bibliographystyle{naturemag}
%\bibliography{active_particles.bib}

\setlength{\bibsep}{0pt plus 0.3ex}
%\todototoc
%\listoftodos

\end{document}